\documentclass[prb, 11pt, aps, showpacs, longbibliography,preprint, endfloats*]{revtex4-2}

\usepackage[dvipsnames]{xcolor}
\usepackage{amssymb,bm}
\usepackage{graphicx}
\usepackage{amsmath} 
\usepackage{setspace}
\usepackage{enumerate}
\usepackage{bbold}
\usepackage{esint}
\usepackage{float}
\usepackage{euscript}
\usepackage[bookmarks, colorlinks=true, breaklinks]{hyperref}
\hypersetup{linkcolor=blue,citecolor=blue,filecolor=black,urlcolor=blue}
\usepackage[normalem]{ulem}

\newcommand{\norm}[1]{\left| {#1} \right|}

\begin{document}

\title{Interplay of Intersite Charge Transfer, Antiferromagnetism, and Strain in Barocaloric ACu$_3$Fe$_4$O$_{12}$ Quadruple 
Perovskites}

\author{J. Delgado-Quesada}
\affiliation{$^{1}$Centro de Investigaci\'{o}n en Ciencia e Ingenier\'{i}a de Materiales, Universidad de Costa Rica, Costa Rica}
\affiliation{$^{2}$Escuela de F\'{i}sica, Universidad de Costa Rica, Costa Rica}
\author{G. G. Guzm\'{a}n-Verri\footnote{gian.guzman@ucr.ac.cr}}
\affiliation{$^{1}$Centro de Investigaci\'{o}n en Ciencia e Ingenier\'{i}a de Materiales, Universidad de Costa Rica, Costa Rica}
\affiliation{$^{2}$Escuela de F\'{i}sica, Universidad de Costa Rica, Costa Rica}

\date{\today}


\begin{abstract}
We develop a minimal Landau theory for the concomitant intersite charge-transfer, antiferromagnetic, and isostructural phase transitions in ACu$_3$Fe$_4$O$_{12}$ perovskites (A = La, Pr, Nd, Sm, Eu, Gd, Tb). The model incorporates the difference in average ligand-hole occupancy between Cu and Fe, the staggered magnetization of the Fe sublattice, volume strain, and intrinsic thermal expansion, together with their couplings. It qualitatively reproduces key thermodynamic properties of the ACu$_3$Fe$_4$O$_{12}$ family, including the staggered magnetization, lattice volume, magnetic susceptibility, and the nearly linear temperature–pressure phase boundary. The framework predicts a pronounced elastic softening near the phase boundary, consistent with experiments where the bulk modulus of the low-pressure, charge-transferred antiferromagnetic phase exceeds that of the high-pressure, non-transferred paramagnetic phase.
It also yields pressure-driven isothermal entropy changes, revealing that the intrinsic thermal expansion of the high- and low-temperature phases significantly shapes the overall barocaloric response. These results contrast with previous analyses of NdCu$_3$Fe$_4$O$_{12}$, where thermal expansion was neglected in the entropy construction, and call for a reevaluation of barocaloric effects in quadruple perovskites. 
\end{abstract}
\maketitle

\section{Introduction}
\vspace{-0.25cm}
Quadruple perovskites (QPs) are complex oxides with the chemical formula AA$^\prime_3$B$_4$O$_{12}$, featuring ordered arrangements of multiple cations on the A and B sites of the perovskite lattice~\cite{Ding2024a, Shimakawa2014a}. Their unique crystal structures enable a wide range of physical and functional phenomena, including large magnetoresistance~\cite{Zeng1999a, Takata2007a}, giant dielectricity~\cite{Subramanian2000a, Homes2001a},  magnetoelectricity~\cite{Zhou2017a}, and catalysis~\cite{Yagi2015a}.

\begin{figure}
    \centering
    \includegraphics[width=0.75\linewidth]{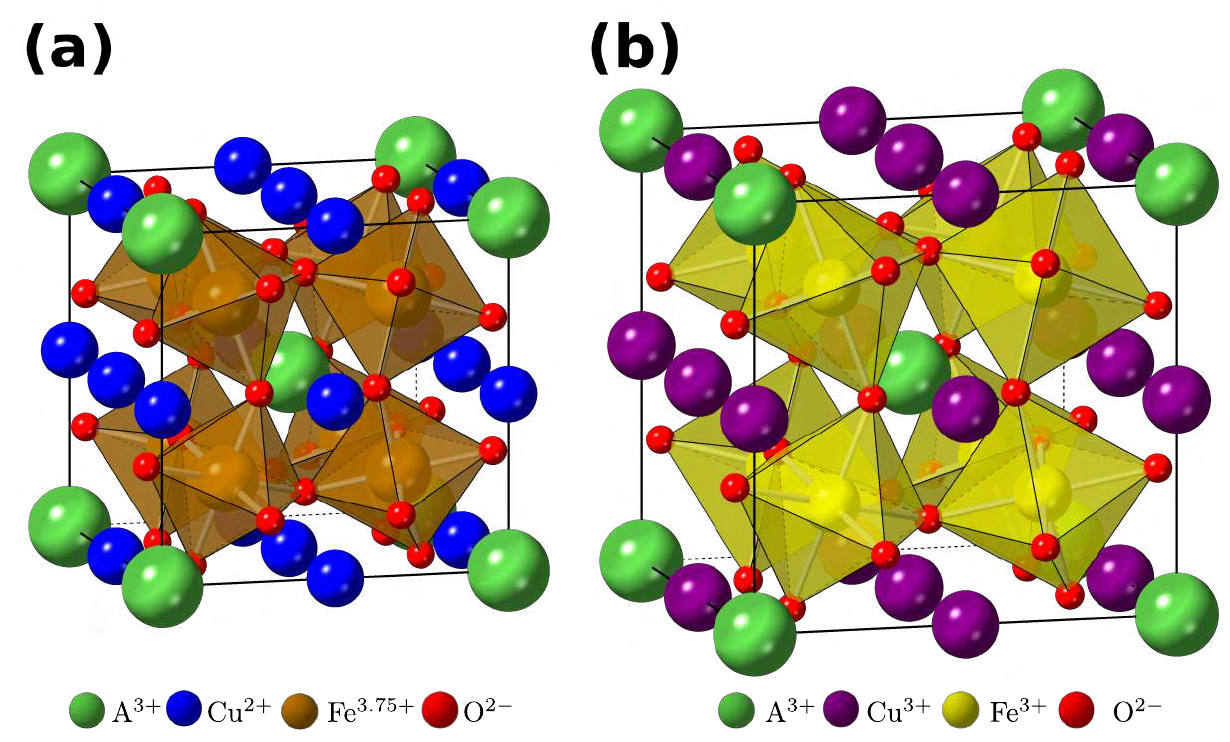}
    \caption{Crystal structure of ACFO  in the (a) high-temperature and (b) low-temperature phases, both adopting  $Im\overline{3}$ symmetry. The change in volume is not to scale and shown schematically for clarity. Generated using CrystalMaker\textsuperscript{\textregistered}.}
    \label{fig:unit_cell}
\end{figure}

In ACu$_3$Fe$_4$O$_{12}$ (ACFO)~\cite{Long2009a, Yamada2013a, Shimakawa2015a, Briere2016a}, where the A-site is occupied by a trivalent lanthanide (La, Pr, Nd, Sm, Eu, Gd, Tb), the high-temperature (HT) phase crystallizes into a cubic $Im\overline{3}$ lattice structure with the nominal ionic formula
A$^{3+}$Cu$^{2+}_3$Fe$^{3.75+}_4$O$_{12}$~(Fig.~\ref{fig:unit_cell}a). Upon cooling, ACFO undergoes an intersite charge transfer (ICT) phase transition in which the A$^\prime$-site Cu$^{2+}$ oxidizes to Cu$^{3+}$, releasing electrons that are transferred to the B-site Fe cations, thereby reducing Fe$^{3.75+}$ to Fe$^{3+}$. The nominal valence state in the low-temperature (LT) becomes A$^{3+}$Cu$^{3+}_3$Fe$^{3+}_4$O$_{12}$,  and the $Im\overline{3}$ space group is retained after an abrupt volume expansion ($\sim$1-2\,\%) at the phase transition (Fig.~\ref{fig:unit_cell}b). Notably, the ICT phase transition in ACFO is concomitant with a first order metal-to-insulator and paramagnetic-to-antiferromagnetic phase transition.  The antiferromagnetic ordering is of the G-type, in which  each magnetic moment of the B-site Fe$^{3+}$ aligns antiparallel to its six nearest iron neighbors. The G-type ordered moments are aligned with one of the cube axes.
No magnetic contribution is observed from the A-site trivalent lanthanide or the A$^\prime$-site Cu$^{+3}$.

It was proposed~\cite{Long2009a} and later confirmed by first-principles calculations~\cite{Li2012a} that the ICT between A$^\prime$-site Cu and B-site Fe is mediated by ligand holes arising from covalent hybridization between Fe/Cu $3d$ and O $2p$ orbitals, rather than by direct electron transfer. In this picture, the ICT process is written as $3d^9 + 4d^5L^{0.75} \rightarrow 3d^9L + 4d^5$, where $L$ denotes a ligand hole. In the HT phase, ligand holes from Fe–O bonds are itinerant, whereas in the LT phase they localize on Cu–O bonds~\cite{Chen2012a}. Because the metal-to-insulator transition is driven by localization of an odd number of itinerant ligand holes without breaking cubic symmetry, it is classified as a Mott transition of the ligand holes~\cite{Chen2012a}.

Isovalent A-site cation substitution across the lanthanide series provides control over the transition temperature $T_c$ in ACFO, which varies from $\sim$360\,K to $\sim$240\,K as the lanthanide size decreases~\cite{Yamada2013a}. Notably, the transition entropies are large ($\sim$60-84\,${\rm J\,K^{-1}\,kg^{-1}}$), making QPs attractive for solid-state thermal energy storage and thermal energy conversion~\cite{Uchimura2020a, Kosugi2021a}. Indeed, nominally reversible~\cite{Moya2020a} thermal changes driven by hydrostatic pressure~\cite{Manosa2013a, Moya2014a, Ichiro2015a} have been quasi-directly determined in NdCu$_3$Fe$_4$O$_{12}$ ($T_c \simeq 310\,{\rm K}$)~\cite{Kosugi2021a} and in related  compounds~\cite{Kosugi2021b, Chen2023a}. Such barocaloric (BC) effects~\cite{Boldrin2021a, Lloveras2023a} are parametrized by isothermal changes in entropy and adiabatic  changes in temperature, which for  NdCu$_3$Fe$_4$O$_{12}$ peak at $\sim$65\,${\rm J K^{-1} kg^{-1}}$ and $\sim$14\,K under $\sim$0.5 GPa at $\sim$294\,K. These values are comparable to the largest changes in inorganic solid materials~\cite{Lloveras2021a}.

The ICT phase transition in ACFO has been previously studied using first-principles calculations~\cite{Li2012a, Rezaei2014a, Meng2017a} and mean-field approximations of the Falikov–Kimball model~\cite{Allub2012a}. While these works provide microscopic insight into the ICT mechanism, the first-principles studies are limited to zero temperature, and the Falikov–Kimball approach neglects the electron-lattice interactions.  Here, we develop a minimal Landau theory that incorporates the coupling between electronic, magnetic, and strain degrees of freedom (DoF)  in order to capture the temperature and pressure dependence of the characteristic thermodynamic features near the ICT phase transition across the ACFO family. We then use this framework to calculate the bulk modulus and BC effects, and to compare our predictions with experimental data.

\section{Model}

\subsection{Order parameters}

We define a non-symmetry breaking electronic order parameter $\Delta \nu$ as the difference in the average ligand-hole occupancy per transition metal ion, i.e.,
\begin{align*}
    \Delta \nu = \nu_{L}(\text{Cu}) -  \nu_{L}(\text{Fe}),    
\end{align*}
where $\nu_{L}(\text{Cu})$ and $\nu_{L}(\text{Fe})$ are the average ligand-hole occupancy per Cu and Fe ion, respectively. Above $T_c$, $\Delta \nu < 0$, since the itinerant ligand holes are primarily concentrated on the Fe ions ($\nu_L(\text{Cu}) \simeq 0$, $\nu_L(\text{Fe}) \simeq 0.75$), whereas below $T_c$, $\Delta \nu > 0$, as the ligand holes localize on the Cu ions ($\nu_L(\text{Cu}) \simeq 1$, $\nu_L(\text{Fe}) \simeq 0$)~\cite{Chen2012a}. To describe the G-type antiferromagnetic order, we use a Néel order parameter ${\bm n}$. We also introduce a ferromagnetic magnetization ${\bm m}$ as an auxiliary order parameter to compute the ferromagnetic susceptibility. Lastly, we account for the elastic volume strain $\eta$, which captures lattice deformations arising from the ICT phase transition, antiferromagnetic ordering, thermal expansion, and applied hydrostatic pressure.

\subsection{Free Energy Density}

We consider an isotropic solid subjected to an applied hydrostatic pressure $P$, where $P=0$ corresponds to ambient pressure. The Landau free energy density $G$ must include both even and odd terms in $\Delta \nu$ as the Cu and Fe sublattices are not crystallographically equivalent. In addition, $G$ must be an even functional of ${\bm n}$ and ${\bm m}$, as it must remain invariant under sublattice exchange symmetry of the Fe ions. Based on these symmetry constraints, we decompose the free-energy density into electronic, magnetic, elastic, and coupling contributions as follows,
\begin{multline}
    \label{eq:potential}
    G=G_0+G_{\rm charge} + G_{\rm magnetic} +G_{\rm strain} \\ +   G_\text{charge-strain} +  G_\text{charge-magnetic}+ G_\text{magnetic-strain}  + P \, \eta,
\end{multline}
where $G_0$ is the free-energy density of the background, $G_{\rm charge}$ is the ICT free-energy density~\cite{Azzolina2020a},
\begin{align*}
    G_{\rm charge}= a_1 \Delta \nu + \frac{a_2}{2} \Delta \nu^2 + \frac{a_3}{3} \Delta \nu^3 + \frac{a_4}{4} \Delta \nu^4,
\end{align*}
$G_{\rm magnetic}$ is the magnetic free-energy density~\cite{Okada1969a},
\begin{align*}
    G_{\rm magnetic}=\frac{b_2}{2} n^2+ \frac{b_2+\lambda}{2} m^2 + \frac{b_4}{4} \left( n^4 + m^4  + 6 n^2  m^2  \right),
\end{align*}
with  $n=\norm{\bm n}$ and $m=\norm{\bm m}$. $G_{\rm strain}$ is the elastic strain free-energy density, which includes thermal expansion~\cite{LandauElasticity},
\begin{align*}
G_{\rm strain}=-B\alpha(T-T_0^\prime) \eta +\frac{B}{2} \eta ^2,
\end{align*}
$G_\text{charge-magnetic}$ is the magnetization-electron coupling energy,
\begin{align*}
    G_{\rm charge\text{-}magnetic} = - \Delta \nu\left(g_{1n} \, n^2 + g_{1m} \, m^2\right),
\end{align*}
$G_\text{charge-strain}$ is a bilinear  charge-strain coupling energy,
\begin{align*}
    G_\text{charge-strain} = - g_2 \, \eta \, \Delta \nu,
\end{align*}
and $G_\text{magnetic-strain}$ is the magnetostrictive coupling energy, 
\begin{align*}
     G_\text{magnetic-strain}= - \eta\left(g_{3n} \, n^2+ g_{3m} \, m^2\right). 
\end{align*}
Here, $a_1 = a_{10}(T - T_\nu)$ and $b_2 = b_{20}(T - T_n)$, where $a_{10}, b_{20}, T_\nu, T_n, a_2, a_3, a_4, b_4, g_1, g_{2n/m}, g_{3n/m}$, and $\lambda$ are model parameters independent of temperature and pressure; $\alpha$ is the bare coefficient of volumetric expansion, and $B$ is the bare bulk modulus. We require $a_{10} > 0$, $a_4 > 0$, $b_{20} > 0$, and $b_4 > 0$ for $G$ to be bounded from below.  $T_0^\prime$ is a reference temperature at which the undeformed state of the solid is defined~\cite{LandauElasticity}.

At thermodynamic equilibrium, the solid relaxes to a stress-free state such that $\left(\partial G / \partial \eta\right)_{n,m} =0$~\cite{Corrales2017a, Marin-Delgado2024a}, which gives the result,
\begin{align}
    \label{eq:eta}
    \eta = \alpha(T-T_0^\prime) - \frac{P}{B} + \frac{g_2}{B} \Delta \nu+ \frac{g_{3n}}{B} n^2 + \frac{g_{3m}}{B}m^2.
\end{align}
Substitution of Eq.~(\ref{eq:eta}) into
Eq.~(\ref{eq:potential}), gives the effective free-energy density,
\begin{align}
    \label{eq:effectiveG}
   \tilde{G}=G_0+\tilde{G}_{\rm charge} + \tilde{G}_{\rm magnetic}  +  \tilde{G}_\text{charge-magnetic} - \frac{[P-B\alpha (T-T_0')]^2}{2B},
\end{align}
where,
\begin{align*}
    \tilde{G}_{\rm charge}= \tilde{a}_1 \Delta \nu + \frac{\tilde{a}_2}{2} \Delta \nu^2 + \frac{a_3}{3} \Delta \nu^3 + \frac{a_4}{4} \Delta \nu^4,
\end{align*}
\begin{align*}
   \tilde{G}_{\rm magnetic} = \frac{\tilde b_{2n}}{2} n^2 + \frac{\tilde b_{2m}+\lambda}{2} m^2
    +\frac{1}{4} \left(\tilde b_{4n}n^4+\tilde b_{4m}m^4 + 6 \, \tilde{b}_{4nm} n^2m^2 \right),
\end{align*}
and,
\begin{align*}
    \tilde{G}_\text{charge-magnetic} = - \Delta \nu\left(\tilde g_{1n}n^2 + \tilde g_{1m}m^2\right),
\end{align*}
with renormalized model parameters $
    \tilde a_1(T, P) =  \tilde a_{10}(T-\tilde T_\nu) + g_2 P/B $, $\tilde a_{10} = a_{10} - g_2\alpha$, $
    \tilde T_\nu = T_\nu + (g_2\alpha/\tilde a_{10})(T_\nu-T_0')$,  $
    \tilde a_2 = a_2 - g_2^2/B$,  $
    \tilde b_{2n}(T, P) =\tilde b_{20n}(T-\tilde T_{n}) + 2g_{3n}P/B$, $ 
    \tilde b_{20n} = b_{20} - 2g_{3n}\alpha$, $\tilde T_{n} = T_n + (2g_{3n}\alpha/ \tilde{b}_{20n})(T_n-T_0')$, $\tilde T_{m} = T_n + (2g_{3m}\alpha/ \tilde{b}_{20m})(T_n-T_0')$, $ 
    \tilde b_{4n} = b_4 - 2g_{3n}^2/B$, $\tilde{b}_{4nm} = (\tilde{b}_{4n}+\tilde{b}_{4m})/2 +\left(g_{3n}^2+g_{3m}^2-2g_{3n}g_{3m}/3\right)/B$, and   
    $\tilde g_{1n} = g_{1n} + g_2\, g_{3n}/B$. $\tilde{b}_{20m}$, $\tilde{b}_{2m}$, $\tilde{b}_{4m}$, and $\tilde{g}_{1m}$ are obtained by replacing $n \to m$ in $\tilde{b}_{20n}$, $\tilde{b}_{2n}$, $\tilde{b}_{4n}$, and $\tilde{g}_{1n}$.

Eq.~(\ref{eq:effectiveG}) is the starting point of our analysis from which we compute the equilibrium configurations $\Delta \nu(T,P)$, $n(T,P)$, and $m(T,P)=0$ as follows, 
\begin{subequations}
    \label{eq:dG}
\begin{align}
    \label{eq:dG1}
  \left( \tilde{b}_{2n} + \tilde{b}_{4n} n^2 - 2\tilde{g}_{1n} \Delta \nu \right)n =0,\\ 
  \label{eq:dG2}
  \tilde a_1 + \tilde a_2 \Delta \nu + a_3 \Delta \nu^2  + a_4 \Delta \nu^3  - \tilde{g}_{1n} n^2 = 0,
\end{align}
\end{subequations}
as well as the relative volume,
\begin{align}
    \label{eq:volume}
    V(T,P)  =  V_0 + \alpha(T-T_0')-\frac{P}{B}+\frac{g_2}{B} \Delta \nu + \frac{g_{3n}}{B}n^2,
\end{align}
the entropy density,
\begin{align}
    \label{eq:entropy}
    S(T,P) =  S_0 + \alpha^2B(T-T_0') - \alpha P - \tilde a_{10}
   \Delta \nu - \frac{\tilde{b}_{20n}}{2} n^2,
\end{align}
the inverse ferromagnetic susceptibility,
\begin{align}
    \label{eq:suscptibility}
    \chi^{-1}(T,P)  =  \tilde b_{2m} + \lambda -  2\tilde g_{1m} \Delta \nu + 3\tilde b_{4nm}n^2,
\end{align}
and the bulk modulus~\cite{Slonczewski1970a},
\begin{multline}
    \label{eq:bulk_modulus}
 B^\prime(T,P) =  B  \\ - \frac{g_2^2(b_2 + 3b_4 n^2-2g_{1n}\Delta\nu-2g_{3n}\eta)+4g_{1n} g_2 g_{3n}n^2+4g_{3n}^2n^2(a_2 + 2a_3\Delta\nu + 3a_4\Delta\nu^2)}{(a_2 + 2a_3\Delta\nu + 3a_4\Delta\nu^2)(b_2 + 3b_4n^2-2g_{1n}\Delta\nu-2g_{3n}\eta)-4g_{1n}^2n^2}.
\end{multline}
Here, $\eta$ is given by Eq.~(\ref{eq:eta}) with $m=0$,  and
$\Delta \nu$ and $n$ are given by Eq.~(\ref{eq:dG}). $V_0 = ( \partial G_0 / \partial P )_T$ and $S_0 =-( \partial G_0 / \partial T )_P$ are the relative volume and entropy density associated with the background, respectively.  

Following the standard convention~\cite{Lloveras2023a}, we calculate isothermal changes in entropy on first compression as follows, \begin{align*}
    \Delta S(T,0 \to P) &=  S(T,P)-S(T,0) 
    =\Delta S_\text{thermal expansion}
    +\Delta S_\text{charge} + \Delta S_\text{magnetic},
   \end{align*}
where,
\begin{align*}
    \Delta S_\text{thermal expansion} &= - \alpha P, \\ 
    \Delta S_\text{charge} &= \tilde{a}_{10} \left[ \Delta \nu(T,0)- \Delta \nu(T,P) \right], \\ \Delta S_\text{magnetic} &=\frac{\tilde{b}_{20n}}{2}\left[ n^2(T,0) - n^2(T,P) \right].
\end{align*}
 $\Delta S_\text{thermal expansion}$, $\Delta S_\text{charge}$, and $\Delta S_\text{magnetic}$ denote the isothermal entropy changes upon first compression, which we associate with the bare thermal expansion, charge, and magnetic DoF, respectively. 
 
\section{Results and discussion}

\begin{figure}[H]
    \centering
    \includegraphics[width=\linewidth]{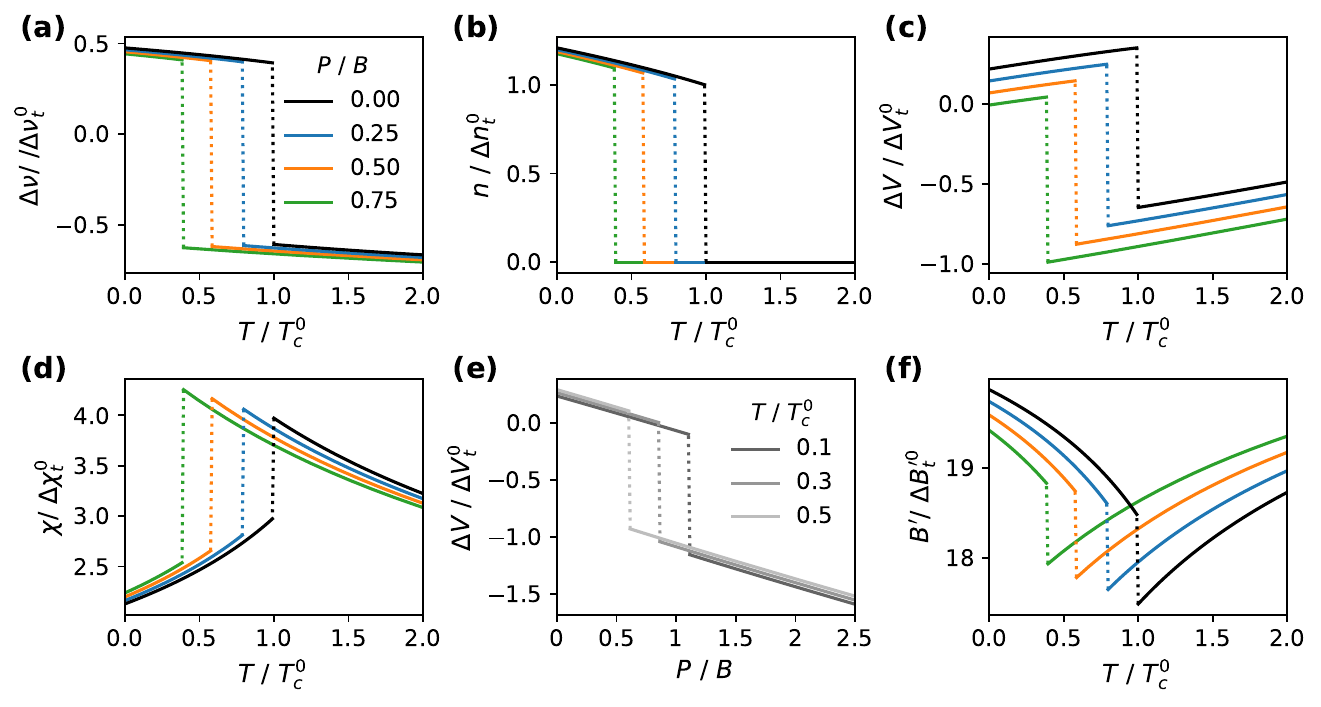}
    \caption{Temperature dependence of (a) the difference in the average ligand-hole occupancy between Cu and Fe, (b) the Néel order parameter, (c) the volume, and (d) the ferromagnetic susceptibility along several isobars. (e) Pressure dependence of the volume along selected isotherms. (f) Temperature dependence of the bulk modulus along several isobars. 
    $\Delta \nu_t^0$, $\Delta n_t^0$, $\Delta V_t^0$, $\Delta \chi_t^0$, and  $\Delta {B'}_t^0$ are the step discontinuities of the ligand-hole imbalance, the Néel order parameter, the volume, the magnetic susceptibility and the bulk modulus at the zero-pressure phase transition, respectively.
    The color scheme in (b), (c), (d), and (f) follows the same pressure assignments as the legend in (a).
    Here and throughout the present work: $a_{10}\,T_\nu/B=2$, $a_2/B=0.5$, $a_3/B=1$, $a_4/B=0.5$, $b_4\,B/(b_{20}^2\,T_\nu^2)=1$, $g_{1n}/(b_{20}\,T_\nu)=1$, $g_{1m}/(b_{20}\,T_\nu)=1$, $g_2/B=1$, 
    $g_{3n}=0$, $g_{3m}=0$, $\alpha \, T_\nu=1$, $T_n/T_\nu=1$, $\lambda/(b_{20}\,T_\nu)=1$, and $T_0^\prime/T_\nu=1$. This parametrization gives $T_c^0/ T_\nu \simeq 0.84$.}
 \label{fig:thermal_properties}
\end{figure}

In ACFO, the relevant physical scenario corresponds to a HT paramagnetic, non–charge-transferred phase; a LT antiferromagnetic, charge-transferred phase; negative thermal expansion at the phase transition; and a transition temperature that decreases under applied pressure. One representative set of model parameters reproducing this behavior is provided in the caption of Fig.~\ref{fig:thermal_properties}. Such a parametrization is used throughout the present work.

At zero pressure, the model exhibits a first-order phase transition at a critical temperature $T^0_c$. This transition is marked by abrupt changes in the ligand-hole imbalance $\Delta \nu$ (Fig.~\ref{fig:thermal_properties}a), N\'eel order parameter $n$ (Fig.~\ref{fig:thermal_properties}b), relative volume change $\Delta V = V-V_0$ (Fig.~\ref{fig:thermal_properties}c), and ferromagnetic susceptibility $\chi$ (Fig.~\ref{fig:thermal_properties}d).
These results are in qualitative agreement with experiments at ambient pressure~\cite{Long2009a, Chen2010a, Yamada2013a}.

\begin{figure}
    \centering
    \includegraphics[width=0.75\linewidth]{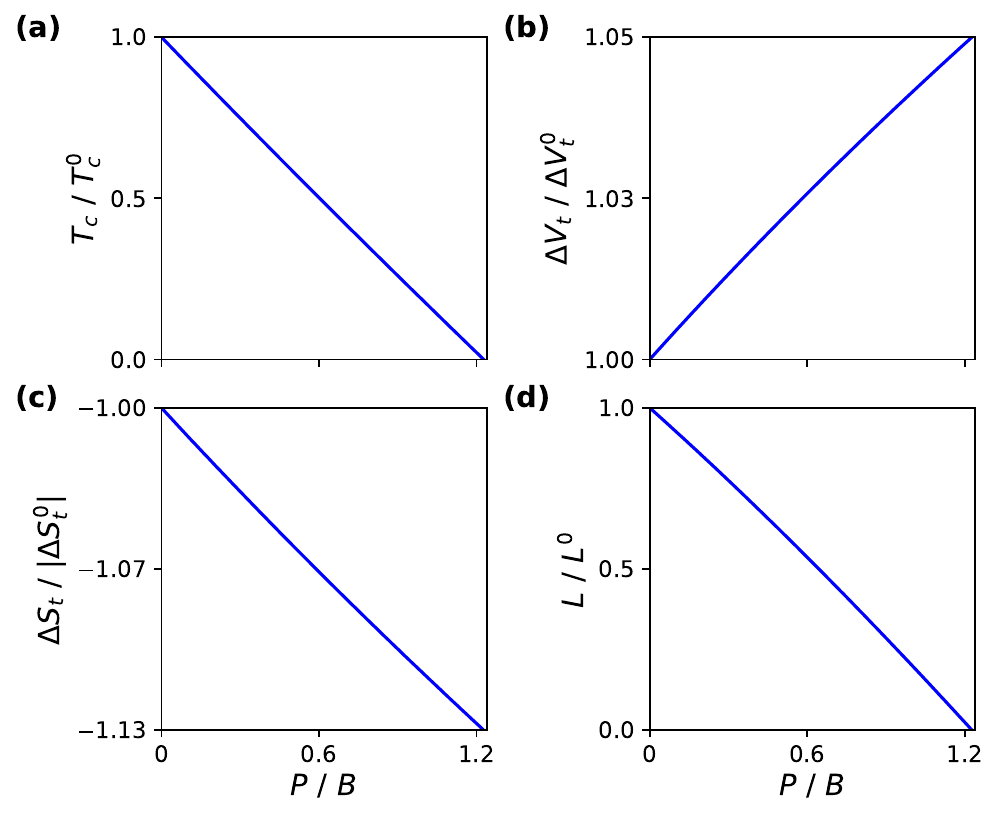}
    \caption{(a) Temperature-pressure  phase boundary of between the LT and HT phases. Pressure dependence of (b) the change $\Delta V_t$ in the relative volume at the phase transition and (c) the change $\Delta S_t$ in the entropy density  at the phase transition, and (d) the latent heat $L=T_c |\Delta S_t|$. $\Delta S_t^0$ and $L^0=T_c^0 |\Delta S_t^0|$ are  the entropy change and the latent heat at the zero-pressure phase transition, respectively.}
    \label{fig:pressure_dependence}
\end{figure}

Hydrostatic pressure leads to a suppression of the transition temperature, consistent with experimental trends~\cite{Long2012a, Kosugi2021a}. As shown in Figs.~\ref{fig:thermal_properties}a-d and~\ref{fig:pressure_dependence}a, increasing pressure shifts the transition to lower temperatures and enhances the step discontinuities in ligand-hole imbalance, staggered magnetization, and relative volume. Figure~\ref{fig:thermal_properties}e shows that, at fixed temperature, the system undergoes a sharp volume collapse at a critical pressure corresponding to the pressure-driven phase transition. This collapse becomes more pronounced  with increasing pressure, as shown in Fig.~\ref{fig:pressure_dependence}b.
 The calculated $V$-$P$ isoterms shown in Fig~\ref{fig:thermal_properties}e are in qualitative agreement with available experimental data~\cite{Long2012a}.

 Based on the pressure effects experimentally observed in LaCu$_3$Fe$_4$O$_{12}$, Long et al.~\cite{Long2012a} argued that hydrostatic pressure mimics the effects of chemical pressure from decreasing the A-site lanthanide radius, a conclusion later confirmed experimentally by Yamada et al.~\cite{Yamada2013a}. Our model supports this view, as the pressure dependence of $n$, $\Delta V$, and $\chi$ (Fig.~\ref{fig:thermal_properties}) follows the same trends reported for chemical pressure~\cite{Long2009a, Kosugi2021a, Yamada2013a}. Comparison with structural data~\cite{Yamada2013a} further suggests that hydrostatic pressure in the HT phase increases the Fe-O-Fe angle, shrinks the Fe-O and A-O bond lengths, and induces little to no variation in the Cu-O bond length.

We now turn to the elastic properties predicted by the model. Figure~\ref{fig:thermal_properties}f shows the temperature dependence of the bulk modulus $B^\prime$ for several isobars, given by Eq.(\ref{eq:bulk_modulus}). The model predicts pronounced elastic softening near the phase boundary, arising from bilinear coupling between charge and strain and linear–quadratic coupling between charge and the staggered magnetization. Notably, $B^\prime$ increases with pressure in the HT phase and decreases in the LT phase. This behavior accounts for the experimental observation that, when the phase transition is pressure-induced, the low-pressure, charge-transferred antiferromagnetic phase has a higher bulk modulus than the high-pressure, non-transferred paramagnetic phase~\cite{Long2012a}.

\begin{figure}[H]
    \centering
    \includegraphics[width=\linewidth]{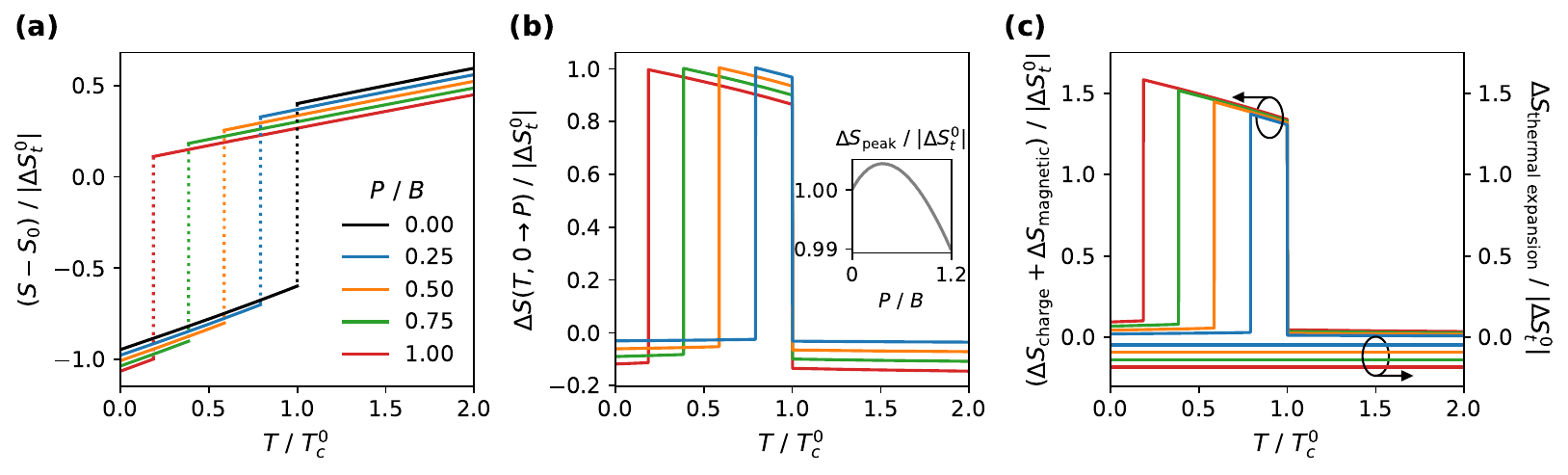}
    \caption{(a) Temperature dependence of the entropy at several isobars. (b) Isothermal entropy changes upon first compression. (c) Isothermal entropy changes upon first compression, showing the combined contribution from charge and magnetic degrees of freedom, and separately, the contribution from bare thermal expansion. The inset in (b) shows the peak values of isothermal changes in entropy during a first compression. The color scheme in (b) and (c) follows the same pressure assignments as the legend in (a).}
    \label{fig:BCE}
\end{figure}

Figure~\ref{fig:BCE}a shows the entropy $S(T, P)$ across a range of temperatures and pressures. The step discontinuity $\Delta S_t$ at the phase transition increases with pressure (Fig.~\ref{fig:pressure_dependence}b), while the associated latent heat $L = T_c |\Delta S_t|$ decreases (Fig.~\ref{fig:pressure_dependence}d) due to the suppression of $T_c$ (Fig.~\ref{fig:pressure_dependence}a). The corresponding isothermal entropy changes upon first compression, $\Delta S(T, 0 \to P)$, are plotted in Fig.~\ref{fig:BCE}b, with peak values $\Delta S_\text{peak}$ shown in the inset. Entropic contributions from the charge ($\Delta S_\text{charge}$) and magnetic ($\Delta S_\text{magnetic}$) degrees of freedom yield inverse BC effects, whereas bare thermal expansion ($\Delta S_\text{thermal expansion}$) produces conventional BC effects (Fig.~\ref{fig:BCE}c). The non-monotonic $\Delta S_\text{peak}$ reflects competition between the positive entropy gain from $\Delta S_\text{charge}$ and $\Delta S_\text{magnetic}$ and the negative 
entropy change $\Delta S_\text{thermal expansion}$: the former dominates for $P/B \lesssim 0.37$, while at higher pressures the latter grows more rapidly, producing a maximum where the two balance.

The predicted isothermal entropy changes (Fig.~\ref{fig:BCE}b) share two key features with NCFO data obtained from the quasi-direct method~\cite{Kosugi2021a}: they are inverse, and their peak values increase with pressure over the measured range. Away from the phase transition, however, the predicted $\Delta S$ remain finite, whereas the reported values drop to zero. We attribute this discrepancy to the omission of the thermal expansion of the individual HT and LT phases in constructing the NCFO entropy function.

\section{Conclusions}

We have presented a Landau theory that captures the interplay of intersite charge transfer, antiferromagnetism, and lattice strain in ACu$_3$Fe$4$O$_{12}$ quadruple perovskites, reproducing the suppression of the transition temperature with pressure, the first-order character of the transition, and the change in volume, magnetization, and entropy.

A central prediction of our theory is a pronounced elastic softening near the temperature-pressure phase boundary, arising from a bilinear coupling between charge and strain and a linear-quadratic coupling between charge and the staggered magnetization. This mechanism accounts for the pressure-induced softening of the bulk modulus in LaCu$_3$Fe$_4$O$_{12}$. The theory could be further tested by measuring the temperature dependence of the bulk modulus under constant pressure~\cite{Mizzi2024a}.

Our model predicts main features of the inverse barocaloric response in ACu$_3$Fe$_4$O$_{12}$, namely, a pressure enhancement of the entropy discontinuity at the phase transition, the corresponding suppression of the latent heat, and the non-monotonic pressure dependence of the peak isothermal entropy change. These trends arise from the competition between the positive entropy gain from the charge–magnetic transition and the negative contribution from bare thermal expansion. These results contrast with previous analyses of NdCu$_3$Fe$_4$O$_{12}$, where thermal expansion was not included in the entropy construction, and call for a reevaluation of barocaloric effects in quadrupole perovskites.

Beyond ACu$_3$Fe$_4$O$_{12}$, our theoretical framework is readily adaptable to model coupled intersite charge-transfer, magnetic, and structural phase transitions in other correlated oxides~\cite{Azuma2007a,Takele2004a,Liu2020a}, offering guiding principles for tuning their electronic, magnetic, elastic, and barocaloric properties for solid-state cooling.

\section{Acknowledgements}

We thank X. Moya and M. Dilshad for discussions on the inverse barocaloric effect. We acknowledge support from the Vice-rectory for Research at the University of Costa Rica under project number C5601.

\newpage


\begin{thebibliography}{40}%
\makeatletter
\providecommand \@ifxundefined [1]{%
 \@ifx{#1\undefined}
}%
\providecommand \@ifnum [1]{%
 \ifnum #1\expandafter \@firstoftwo
 \else \expandafter \@secondoftwo
 \fi
}%
\providecommand \@ifx [1]{%
 \ifx #1\expandafter \@firstoftwo
 \else \expandafter \@secondoftwo
 \fi
}%
\providecommand \natexlab [1]{#1}%
\providecommand \enquote  [1]{``#1''}%
\providecommand \bibnamefont  [1]{#1}%
\providecommand \bibfnamefont [1]{#1}%
\providecommand \citenamefont [1]{#1}%
\providecommand \href@noop [0]{\@secondoftwo}%
\providecommand \href [0]{\begingroup \@sanitize@url \@href}%
\providecommand \@href[1]{\@@startlink{#1}\@@href}%
\providecommand \@@href[1]{\endgroup#1\@@endlink}%
\providecommand \@sanitize@url [0]{\catcode `\\12\catcode `\$12\catcode `\&12\catcode `\#12\catcode `\^12\catcode `\_12\catcode `\%12\relax}%
\providecommand \@@startlink[1]{}%
\providecommand \@@endlink[0]{}%
\providecommand \url  [0]{\begingroup\@sanitize@url \@url }%
\providecommand \@url [1]{\endgroup\@href {#1}{\urlprefix }}%
\providecommand \urlprefix  [0]{URL }%
\providecommand \Eprint [0]{\href }%
\providecommand \doibase [0]{https://doi.org/}%
\providecommand \selectlanguage [0]{\@gobble}%
\providecommand \bibinfo  [0]{\@secondoftwo}%
\providecommand \bibfield  [0]{\@secondoftwo}%
\providecommand \translation [1]{[#1]}%
\providecommand \BibitemOpen [0]{}%
\providecommand \bibitemStop [0]{}%
\providecommand \bibitemNoStop [0]{.\EOS\space}%
\providecommand \EOS [0]{\spacefactor3000\relax}%
\providecommand \BibitemShut  [1]{\csname bibitem#1\endcsname}%
\let\auto@bib@innerbib\@empty
\bibitem [{\citenamefont {Ding}\ and\ \citenamefont {Zhu}(2024)}]{Ding2024a}%
  \BibitemOpen
  \bibfield  {author} {\bibinfo {author} {\bibfnamefont {J.}~\bibnamefont {Ding}}\ and\ \bibinfo {author} {\bibfnamefont {X.}~\bibnamefont {Zhu}},\ }\bibfield  {title} {\bibinfo {title} {Research progress on quadruple perovskite oxides},\ }\href {https://doi.org/10.1039/D4TC01467G} {\bibfield  {journal} {\bibinfo  {journal} {J. Mater. Chem. C}\ }\textbf {\bibinfo {volume} {12}},\ \bibinfo {pages} {9510} (\bibinfo {year} {2024})}\BibitemShut {NoStop}%
\bibitem [{\citenamefont {Shimakawa}\ and\ \citenamefont {Mizumaki}(2014)}]{Shimakawa2014a}%
  \BibitemOpen
  \bibfield  {author} {\bibinfo {author} {\bibfnamefont {Y.}~\bibnamefont {Shimakawa}}\ and\ \bibinfo {author} {\bibfnamefont {M.}~\bibnamefont {Mizumaki}},\ }\bibfield  {title} {\bibinfo {title} {Multiple magnetic interactions in a-site-ordered perovskite-structure oxides},\ }\href {https://doi.org/10.1088/0953-8984/26/47/473203} {\bibfield  {journal} {\bibinfo  {journal} {J. Phys.: Cond. Matt.}\ }\textbf {\bibinfo {volume} {26}},\ \bibinfo {pages} {473203} (\bibinfo {year} {2014})}\BibitemShut {NoStop}%
\bibitem [{\citenamefont {Zeng}\ \emph {et~al.}(1999)\citenamefont {Zeng}, \citenamefont {Greenblatt}, \citenamefont {Subramanian},\ and\ \citenamefont {Croft}}]{Zeng1999a}%
  \BibitemOpen
  \bibfield  {author} {\bibinfo {author} {\bibfnamefont {Z.}~\bibnamefont {Zeng}}, \bibinfo {author} {\bibfnamefont {M.}~\bibnamefont {Greenblatt}}, \bibinfo {author} {\bibfnamefont {M.~A.}\ \bibnamefont {Subramanian}},\ and\ \bibinfo {author} {\bibfnamefont {M.}~\bibnamefont {Croft}},\ }\bibfield  {title} {\bibinfo {title} {{Large low-field magnetoresistance in perovskite-type CaCu$_3$Mn$_4$O$_{12}$ without double exchange}},\ }\href {https://doi.org/10.1103/PhysRevLett.82.3164} {\bibfield  {journal} {\bibinfo  {journal} {Phys. Rev. Lett.}\ }\textbf {\bibinfo {volume} {82}},\ \bibinfo {pages} {3164} (\bibinfo {year} {1999})}\BibitemShut {NoStop}%
\bibitem [{\citenamefont {Takata}\ \emph {et~al.}(2007)\citenamefont {Takata}, \citenamefont {Yamada}, \citenamefont {Azuma}, \citenamefont {Takano},\ and\ \citenamefont {Shimakawa}}]{Takata2007a}%
  \BibitemOpen
  \bibfield  {author} {\bibinfo {author} {\bibfnamefont {K.}~\bibnamefont {Takata}}, \bibinfo {author} {\bibfnamefont {I.}~\bibnamefont {Yamada}}, \bibinfo {author} {\bibfnamefont {M.}~\bibnamefont {Azuma}}, \bibinfo {author} {\bibfnamefont {M.}~\bibnamefont {Takano}},\ and\ \bibinfo {author} {\bibfnamefont {Y.}~\bibnamefont {Shimakawa}},\ }\bibfield  {title} {\bibinfo {title} {{Magnetoresistance and electronic structure of the half-metallic ferrimagnet $\mathrm{Bi}{\mathrm{Cu}}_{3}{\mathrm{Mn}}_{4}{\mathrm{O}}_{12}$}},\ }\href {https://doi.org/10.1103/PhysRevB.76.024429} {\bibfield  {journal} {\bibinfo  {journal} {Phys. Rev. B}\ }\textbf {\bibinfo {volume} {76}},\ \bibinfo {pages} {024429} (\bibinfo {year} {2007})}\BibitemShut {NoStop}%
\bibitem [{\citenamefont {Subramanian}\ \emph {et~al.}(2000)\citenamefont {Subramanian}, \citenamefont {Li}, \citenamefont {Duan}, \citenamefont {Reisner},\ and\ \citenamefont {Sleight}}]{Subramanian2000a}%
  \BibitemOpen
  \bibfield  {author} {\bibinfo {author} {\bibfnamefont {M.}~\bibnamefont {Subramanian}}, \bibinfo {author} {\bibfnamefont {D.}~\bibnamefont {Li}}, \bibinfo {author} {\bibfnamefont {N.}~\bibnamefont {Duan}}, \bibinfo {author} {\bibfnamefont {B.}~\bibnamefont {Reisner}},\ and\ \bibinfo {author} {\bibfnamefont {A.}~\bibnamefont {Sleight}},\ }\bibfield  {title} {\bibinfo {title} {{High Dielectric Constant in ACu$_3$Ti$_4$O$_{12}$ and ACu$_3$Ti$_4$Fe$_{12}$ Phases}},\ }\href {https://doi.org/https://doi.org/10.1006/jssc.2000.8703} {\bibfield  {journal} {\bibinfo  {journal} {J. Solid State Chem.}\ }\textbf {\bibinfo {volume} {151}},\ \bibinfo {pages} {323} (\bibinfo {year} {2000})}\BibitemShut {NoStop}%
\bibitem [{\citenamefont {Homes}\ \emph {et~al.}(2001)\citenamefont {Homes}, \citenamefont {Vogt}, \citenamefont {Shapiro}, \citenamefont {Wakimoto},\ and\ \citenamefont {Ramirez}}]{Homes2001a}%
  \BibitemOpen
  \bibfield  {author} {\bibinfo {author} {\bibfnamefont {C.~C.}\ \bibnamefont {Homes}}, \bibinfo {author} {\bibfnamefont {T.}~\bibnamefont {Vogt}}, \bibinfo {author} {\bibfnamefont {S.~M.}\ \bibnamefont {Shapiro}}, \bibinfo {author} {\bibfnamefont {S.}~\bibnamefont {Wakimoto}},\ and\ \bibinfo {author} {\bibfnamefont {A.~P.}\ \bibnamefont {Ramirez}},\ }\bibfield  {title} {\bibinfo {title} {Optical response of high-dielectric-constant perovskite-related oxide},\ }\href {https://doi.org/10.1126/science.1061655} {\bibfield  {journal} {\bibinfo  {journal} {Science}\ }\textbf {\bibinfo {volume} {293}},\ \bibinfo {pages} {673} (\bibinfo {year} {2001})}\BibitemShut {NoStop}%
\bibitem [{\citenamefont {Zhou}\ \emph {et~al.}(2017)\citenamefont {Zhou}, \citenamefont {Dai}, \citenamefont {Chai}, \citenamefont {Zhang}, \citenamefont {Dong}, \citenamefont {Cao}, \citenamefont {Calder}, \citenamefont {Yin}, \citenamefont {Wang}, \citenamefont {Shen}, \citenamefont {Liu}, \citenamefont {Saito}, \citenamefont {Shimakawa}, \citenamefont {Hojo}, \citenamefont {Ikuhara}, \citenamefont {Azuma}, \citenamefont {Hu}, \citenamefont {Sun}, \citenamefont {Jin},\ and\ \citenamefont {Long}}]{Zhou2017a}%
  \BibitemOpen
  \bibfield  {author} {\bibinfo {author} {\bibfnamefont {L.}~\bibnamefont {Zhou}}, \bibinfo {author} {\bibfnamefont {J.}~\bibnamefont {Dai}}, \bibinfo {author} {\bibfnamefont {Y.}~\bibnamefont {Chai}}, \bibinfo {author} {\bibfnamefont {H.}~\bibnamefont {Zhang}}, \bibinfo {author} {\bibfnamefont {S.}~\bibnamefont {Dong}}, \bibinfo {author} {\bibfnamefont {H.}~\bibnamefont {Cao}}, \bibinfo {author} {\bibfnamefont {S.}~\bibnamefont {Calder}}, \bibinfo {author} {\bibfnamefont {Y.}~\bibnamefont {Yin}}, \bibinfo {author} {\bibfnamefont {X.}~\bibnamefont {Wang}}, \bibinfo {author} {\bibfnamefont {X.}~\bibnamefont {Shen}}, \bibinfo {author} {\bibfnamefont {Z.}~\bibnamefont {Liu}}, \bibinfo {author} {\bibfnamefont {T.}~\bibnamefont {Saito}}, \bibinfo {author} {\bibfnamefont {Y.}~\bibnamefont {Shimakawa}}, \bibinfo {author} {\bibfnamefont {H.}~\bibnamefont {Hojo}}, \bibinfo {author} {\bibfnamefont {Y.}~\bibnamefont {Ikuhara}}, \bibinfo {author} {\bibfnamefont {M.}~\bibnamefont {Azuma}}, \bibinfo {author} {\bibfnamefont
  {Z.}~\bibnamefont {Hu}}, \bibinfo {author} {\bibfnamefont {Y.}~\bibnamefont {Sun}}, \bibinfo {author} {\bibfnamefont {C.}~\bibnamefont {Jin}},\ and\ \bibinfo {author} {\bibfnamefont {Y.}~\bibnamefont {Long}},\ }\bibfield  {title} {\bibinfo {title} {{Realization of large electric polarization and strong magnetoelectric oupling in BiMn$_3$Cr$_4$O$_{12}$}},\ }\href {https://advanced.onlinelibrary.wiley.com/doi/abs/10.1002/adma.201703435} {\bibfield  {journal} {\bibinfo  {journal} {Adv. Mater.}\ }\textbf {\bibinfo {volume} {29}},\ \bibinfo {pages} {1703435} (\bibinfo {year} {2017})}\BibitemShut {NoStop}%
\bibitem [{\citenamefont {Yagi}\ \emph {et~al.}(2015)\citenamefont {Yagi}, \citenamefont {Yamada}, \citenamefont {Tsukasaki}, \citenamefont {Seno}, \citenamefont {Murakami}, \citenamefont {Fujii}, \citenamefont {Chen}, \citenamefont {Umezawa}, \citenamefont {Abe}, \citenamefont {Nishiyama},\ and\ \citenamefont {Mori}}]{Yagi2015a}%
  \BibitemOpen
  \bibfield  {author} {\bibinfo {author} {\bibfnamefont {S.}~\bibnamefont {Yagi}}, \bibinfo {author} {\bibfnamefont {I.}~\bibnamefont {Yamada}}, \bibinfo {author} {\bibfnamefont {H.}~\bibnamefont {Tsukasaki}}, \bibinfo {author} {\bibfnamefont {A.}~\bibnamefont {Seno}}, \bibinfo {author} {\bibfnamefont {M.}~\bibnamefont {Murakami}}, \bibinfo {author} {\bibfnamefont {H.}~\bibnamefont {Fujii}}, \bibinfo {author} {\bibfnamefont {H.}~\bibnamefont {Chen}}, \bibinfo {author} {\bibfnamefont {N.}~\bibnamefont {Umezawa}}, \bibinfo {author} {\bibfnamefont {H.}~\bibnamefont {Abe}}, \bibinfo {author} {\bibfnamefont {N.}~\bibnamefont {Nishiyama}},\ and\ \bibinfo {author} {\bibfnamefont {S.}~\bibnamefont {Mori}},\ }\bibfield  {title} {\bibinfo {title} {Covalency-reinforced oxygen evolution reaction catalyst},\ }\href {https://doi.org/10.1038/ncomms9249} {\bibfield  {journal} {\bibinfo  {journal} {Nature Commun.}\ }\textbf {\bibinfo {volume} {6}},\ \bibinfo {pages} {8249} (\bibinfo {year} {2015})}\BibitemShut {NoStop}%
\bibitem [{\citenamefont {Long}\ \emph {et~al.}(2009)\citenamefont {Long}, \citenamefont {Hayashi}, \citenamefont {Saito}, \citenamefont {Azuma}, \citenamefont {Muranaka},\ and\ \citenamefont {Shimakawa}}]{Long2009a}%
  \BibitemOpen
  \bibfield  {author} {\bibinfo {author} {\bibfnamefont {Y.~W.}\ \bibnamefont {Long}}, \bibinfo {author} {\bibfnamefont {N.}~\bibnamefont {Hayashi}}, \bibinfo {author} {\bibfnamefont {T.}~\bibnamefont {Saito}}, \bibinfo {author} {\bibfnamefont {M.}~\bibnamefont {Azuma}}, \bibinfo {author} {\bibfnamefont {S.}~\bibnamefont {Muranaka}},\ and\ \bibinfo {author} {\bibfnamefont {Y.}~\bibnamefont {Shimakawa}},\ }\bibfield  {title} {\bibinfo {title} {{Temperature-induced A-B intersite charge transfer in an A-site-ordered LaCu$_3$Fe$_4$O$_{12}$ perovskite}},\ }\href {https://doi.org/10.1038/nature07816} {\bibfield  {journal} {\bibinfo  {journal} {Nature}\ }\textbf {\bibinfo {volume} {458}},\ \bibinfo {pages} {60} (\bibinfo {year} {2009})}\BibitemShut {NoStop}%
\bibitem [{\citenamefont {Yamada}\ \emph {et~al.}(2013)\citenamefont {Yamada}, \citenamefont {Etani}, \citenamefont {Tsuchida}, \citenamefont {Marukawa}, \citenamefont {Hayashi}, \citenamefont {Kawakami}, \citenamefont {Mizumaki}, \citenamefont {Ohgushi}, \citenamefont {Kusano}, \citenamefont {Kim}, \citenamefont {Tsuji}, \citenamefont {Takahashi}, \citenamefont {Nishiyama}, \citenamefont {Inoue}, \citenamefont {Irifune},\ and\ \citenamefont {Takano}}]{Yamada2013a}%
  \BibitemOpen
  \bibfield  {author} {\bibinfo {author} {\bibfnamefont {I.}~\bibnamefont {Yamada}}, \bibinfo {author} {\bibfnamefont {H.}~\bibnamefont {Etani}}, \bibinfo {author} {\bibfnamefont {K.}~\bibnamefont {Tsuchida}}, \bibinfo {author} {\bibfnamefont {S.}~\bibnamefont {Marukawa}}, \bibinfo {author} {\bibfnamefont {N.}~\bibnamefont {Hayashi}}, \bibinfo {author} {\bibfnamefont {T.}~\bibnamefont {Kawakami}}, \bibinfo {author} {\bibfnamefont {M.}~\bibnamefont {Mizumaki}}, \bibinfo {author} {\bibfnamefont {K.}~\bibnamefont {Ohgushi}}, \bibinfo {author} {\bibfnamefont {Y.}~\bibnamefont {Kusano}}, \bibinfo {author} {\bibfnamefont {J.}~\bibnamefont {Kim}}, \bibinfo {author} {\bibfnamefont {N.}~\bibnamefont {Tsuji}}, \bibinfo {author} {\bibfnamefont {R.}~\bibnamefont {Takahashi}}, \bibinfo {author} {\bibfnamefont {N.}~\bibnamefont {Nishiyama}}, \bibinfo {author} {\bibfnamefont {T.}~\bibnamefont {Inoue}}, \bibinfo {author} {\bibfnamefont {T.}~\bibnamefont {Irifune}},\ and\ \bibinfo {author} {\bibfnamefont {M.}~\bibnamefont
  {Takano}},\ }\bibfield  {title} {\bibinfo {title} {Control of bond-strain-induced electronic phase transitions in iron perovskites},\ }\href {https://doi.org/10.1021/ic402344m} {\bibfield  {journal} {\bibinfo  {journal} {Inorg. Chem.}\ }\textbf {\bibinfo {volume} {52}},\ \bibinfo {pages} {13751} (\bibinfo {year} {2013})}\BibitemShut {NoStop}%
\bibitem [{\citenamefont {Shimakawa}(2015)}]{Shimakawa2015a}%
  \BibitemOpen
  \bibfield  {author} {\bibinfo {author} {\bibfnamefont {Y.}~\bibnamefont {Shimakawa}},\ }\bibfield  {title} {\bibinfo {title} {{Crystal and magnetic structures of CaCu$_3$Fe$_4$O$_{12}$ and LaCu$_3$Fe$_4$O$_{12}$: distinct charge transitions of unusual high valence Fe}},\ }\href {https://doi.org/10.1088/0022-3727/48/50/504006} {\bibfield  {journal} {\bibinfo  {journal} {J. Phys. D: Appl. Phys.}\ }\textbf {\bibinfo {volume} {48}},\ \bibinfo {pages} {504006} (\bibinfo {year} {2015})}\BibitemShut {NoStop}%
\bibitem [{\citenamefont {Bri{\`e}re}\ \emph {et~al.}(2016)\citenamefont {Bri{\`e}re}, \citenamefont {Kalinko}, \citenamefont {Yamada}, \citenamefont {Roy}, \citenamefont {Brubach}, \citenamefont {Sopracase}, \citenamefont {Zaghrioui},\ and\ \citenamefont {Phuoc}}]{Briere2016a}%
  \BibitemOpen
  \bibfield  {author} {\bibinfo {author} {\bibfnamefont {B.}~\bibnamefont {Bri{\`e}re}}, \bibinfo {author} {\bibfnamefont {A.}~\bibnamefont {Kalinko}}, \bibinfo {author} {\bibfnamefont {I.}~\bibnamefont {Yamada}}, \bibinfo {author} {\bibfnamefont {P.}~\bibnamefont {Roy}}, \bibinfo {author} {\bibfnamefont {J.~B.}\ \bibnamefont {Brubach}}, \bibinfo {author} {\bibfnamefont {R.}~\bibnamefont {Sopracase}}, \bibinfo {author} {\bibfnamefont {M.}~\bibnamefont {Zaghrioui}},\ and\ \bibinfo {author} {\bibfnamefont {V.~T.}\ \bibnamefont {Phuoc}},\ }\bibfield  {title} {\bibinfo {title} {{On the energy scale involved in the metal to insulator transition of quadruple perovskite EuCu$_3$Fe$_4$O$_{12}$: infrared spectroscopy and ab-initio calculations}},\ }\href {https://doi.org/10.1038/srep28624} {\bibfield  {journal} {\bibinfo  {journal} {Sci. Rep.}\ }\textbf {\bibinfo {volume} {6}},\ \bibinfo {pages} {28624} (\bibinfo {year} {2016})}\BibitemShut {NoStop}%
\bibitem [{\citenamefont {Li}\ \emph {et~al.}(2012)\citenamefont {Li}, \citenamefont {Lv}, \citenamefont {Wang}, \citenamefont {Xia}, \citenamefont {Bai}, \citenamefont {Liu},\ and\ \citenamefont {Meng}}]{Li2012a}%
  \BibitemOpen
  \bibfield  {author} {\bibinfo {author} {\bibfnamefont {H.}~\bibnamefont {Li}}, \bibinfo {author} {\bibfnamefont {S.}~\bibnamefont {Lv}}, \bibinfo {author} {\bibfnamefont {Z.}~\bibnamefont {Wang}}, \bibinfo {author} {\bibfnamefont {Y.}~\bibnamefont {Xia}}, \bibinfo {author} {\bibfnamefont {Y.}~\bibnamefont {Bai}}, \bibinfo {author} {\bibfnamefont {X.}~\bibnamefont {Liu}},\ and\ \bibinfo {author} {\bibfnamefont {J.}~\bibnamefont {Meng}},\ }\bibfield  {title} {\bibinfo {title} {{Mechanism of A-B intersite charge transfer and negative thermal expansion in A-site-ordered perovskite LaCu$_4$Fe$_4$O$_{12}$}},\ }\href {https://doi.org/10.1063/1.4721408} {\bibfield  {journal} {\bibinfo  {journal} {J. Appl. Phys.}\ }\textbf {\bibinfo {volume} {111}},\ \bibinfo {pages} {103718} (\bibinfo {year} {2012})}\BibitemShut {NoStop}%
\bibitem [{\citenamefont {Chen}\ \emph {et~al.}(2012)\citenamefont {Chen}, \citenamefont {Saito}, \citenamefont {Hayashi}, \citenamefont {Takano},\ and\ \citenamefont {Shimakawa}}]{Chen2012a}%
  \BibitemOpen
  \bibfield  {author} {\bibinfo {author} {\bibfnamefont {W.~T.}\ \bibnamefont {Chen}}, \bibinfo {author} {\bibfnamefont {T.}~\bibnamefont {Saito}}, \bibinfo {author} {\bibfnamefont {N.}~\bibnamefont {Hayashi}}, \bibinfo {author} {\bibfnamefont {M.}~\bibnamefont {Takano}},\ and\ \bibinfo {author} {\bibfnamefont {Y.}~\bibnamefont {Shimakawa}},\ }\bibfield  {title} {\bibinfo {title} {{Ligand-hole localization in oxides with unusual valence Fe}},\ }\bibfield  {journal} {\bibinfo  {journal} {Sci. Rep.}\ }\textbf {\bibinfo {volume} {2}},\ \href {https://doi.org/10.1038/srep00449} {10.1038/srep00449} (\bibinfo {year} {2012})\BibitemShut {NoStop}%
\bibitem [{\citenamefont {Uchimura}\ and\ \citenamefont {Yamada}(2020)}]{Uchimura2020a}%
  \BibitemOpen
  \bibfield  {author} {\bibinfo {author} {\bibfnamefont {T.}~\bibnamefont {Uchimura}}\ and\ \bibinfo {author} {\bibfnamefont {I.}~\bibnamefont {Yamada}},\ }\bibfield  {title} {\bibinfo {title} {A robust thermal-energy-storage property associated with electronic phase transitions for quadruple perovskite oxides},\ }\href {https://doi.org/10.1039/D0CC01715A} {\bibfield  {journal} {\bibinfo  {journal} {Chem. Commun.}\ }\textbf {\bibinfo {volume} {56}},\ \bibinfo {pages} {5500} (\bibinfo {year} {2020})}\BibitemShut {NoStop}%
\bibitem [{\citenamefont {Kosugi}\ \emph {et~al.}(2021{\natexlab{a}})\citenamefont {Kosugi}, \citenamefont {Goto}, \citenamefont {Tan}, \citenamefont {Fujita}, \citenamefont {Saito}, \citenamefont {Kamiyama}, \citenamefont {Chen}, \citenamefont {Chuang}, \citenamefont {Sheu}, \citenamefont {Kan},\ and\ \citenamefont {Shimakawa}}]{Kosugi2021a}%
  \BibitemOpen
  \bibfield  {author} {\bibinfo {author} {\bibfnamefont {Y.}~\bibnamefont {Kosugi}}, \bibinfo {author} {\bibfnamefont {M.}~\bibnamefont {Goto}}, \bibinfo {author} {\bibfnamefont {Z.}~\bibnamefont {Tan}}, \bibinfo {author} {\bibfnamefont {A.}~\bibnamefont {Fujita}}, \bibinfo {author} {\bibfnamefont {T.}~\bibnamefont {Saito}}, \bibinfo {author} {\bibfnamefont {T.}~\bibnamefont {Kamiyama}}, \bibinfo {author} {\bibfnamefont {W.~T.}\ \bibnamefont {Chen}}, \bibinfo {author} {\bibfnamefont {Y.~C.}\ \bibnamefont {Chuang}}, \bibinfo {author} {\bibfnamefont {H.~S.}\ \bibnamefont {Sheu}}, \bibinfo {author} {\bibfnamefont {D.}~\bibnamefont {Kan}},\ and\ \bibinfo {author} {\bibfnamefont {Y.}~\bibnamefont {Shimakawa}},\ }\bibfield  {title} {\bibinfo {title} {Colossal barocaloric effect by large latent heat produced by first-order intersite-charge-transfer transition},\ }\href {https://doi.org/10.1002/adfm.202009476} {\bibfield  {journal} {\bibinfo  {journal} {Adv. Func. Mater.}\ }\textbf {\bibinfo {volume} {31}} (\bibinfo
  {year} {2021}{\natexlab{a}})}\BibitemShut {NoStop}%
\bibitem [{\citenamefont {Moya}\ and\ \citenamefont {Mathur}(2020)}]{Moya2020a}%
  \BibitemOpen
  \bibfield  {author} {\bibinfo {author} {\bibfnamefont {X.}~\bibnamefont {Moya}}\ and\ \bibinfo {author} {\bibfnamefont {N.~D.}\ \bibnamefont {Mathur}},\ }\bibfield  {title} {\bibinfo {title} {Caloric materials for cooling and heating},\ }\href {https://doi.org/10.1126/science.abb0973} {\bibfield  {journal} {\bibinfo  {journal} {Science}\ }\textbf {\bibinfo {volume} {370}},\ \bibinfo {pages} {797} (\bibinfo {year} {2020})}\BibitemShut {NoStop}%
\bibitem [{\citenamefont {Mañosa}\ \emph {et~al.}(2013)\citenamefont {Mañosa}, \citenamefont {Planes},\ and\ \citenamefont {Acet}}]{Manosa2013a}%
  \BibitemOpen
  \bibfield  {author} {\bibinfo {author} {\bibfnamefont {L.}~\bibnamefont {Mañosa}}, \bibinfo {author} {\bibfnamefont {A.}~\bibnamefont {Planes}},\ and\ \bibinfo {author} {\bibfnamefont {M.}~\bibnamefont {Acet}},\ }\bibfield  {title} {\bibinfo {title} {Advanced materials for solid-state refrigeration},\ }\href {https://doi.org/10.1039/C3TA01289A} {\bibfield  {journal} {\bibinfo  {journal} {J. Mater. Chem. A}\ }\textbf {\bibinfo {volume} {1}},\ \bibinfo {pages} {4925} (\bibinfo {year} {2013})}\BibitemShut {NoStop}%
\bibitem [{\citenamefont {Moya}\ \emph {et~al.}(2014)\citenamefont {Moya}, \citenamefont {Kar-Narayan},\ and\ \citenamefont {Mathur}}]{Moya2014a}%
  \BibitemOpen
  \bibfield  {author} {\bibinfo {author} {\bibfnamefont {X.}~\bibnamefont {Moya}}, \bibinfo {author} {\bibfnamefont {S.}~\bibnamefont {Kar-Narayan}},\ and\ \bibinfo {author} {\bibfnamefont {N.}~\bibnamefont {Mathur}},\ }\bibfield  {title} {\bibinfo {title} {Caloric materials near ferroic phase transitions},\ }\href {https://doi.org/10.1038/nmat3951} {\bibfield  {journal} {\bibinfo  {journal} {Nature Mater.}\ }\textbf {\bibinfo {volume} {13}},\ \bibinfo {pages} {439} (\bibinfo {year} {2014})}\BibitemShut {NoStop}%
\bibitem [{\citenamefont {Takeuchi}\ and\ \citenamefont {Sandeman}(2015)}]{Ichiro2015a}%
  \BibitemOpen
  \bibfield  {author} {\bibinfo {author} {\bibfnamefont {I.}~\bibnamefont {Takeuchi}}\ and\ \bibinfo {author} {\bibfnamefont {K.}~\bibnamefont {Sandeman}},\ }\bibfield  {title} {\bibinfo {title} {Solid-state cooling with caloric materials},\ }\href {https://doi.org/10.1063/PT.3.3022} {\bibfield  {journal} {\bibinfo  {journal} {Physics Today}\ }\textbf {\bibinfo {volume} {68}},\ \bibinfo {pages} {48} (\bibinfo {year} {2015})}\BibitemShut {NoStop}%
\bibitem [{\citenamefont {Kosugi}\ \emph {et~al.}(2021{\natexlab{b}})\citenamefont {Kosugi}, \citenamefont {Goto}, \citenamefont {Tan}, \citenamefont {Kan}, \citenamefont {Isobe}, \citenamefont {Yoshii}, \citenamefont {Mizumaki}, \citenamefont {Fujita}, \citenamefont {Takagi},\ and\ \citenamefont {Shimakawa}}]{Kosugi2021b}%
  \BibitemOpen
  \bibfield  {author} {\bibinfo {author} {\bibfnamefont {Y.}~\bibnamefont {Kosugi}}, \bibinfo {author} {\bibfnamefont {M.}~\bibnamefont {Goto}}, \bibinfo {author} {\bibfnamefont {Z.}~\bibnamefont {Tan}}, \bibinfo {author} {\bibfnamefont {D.}~\bibnamefont {Kan}}, \bibinfo {author} {\bibfnamefont {M.}~\bibnamefont {Isobe}}, \bibinfo {author} {\bibfnamefont {K.}~\bibnamefont {Yoshii}}, \bibinfo {author} {\bibfnamefont {M.}~\bibnamefont {Mizumaki}}, \bibinfo {author} {\bibfnamefont {A.}~\bibnamefont {Fujita}}, \bibinfo {author} {\bibfnamefont {H.}~\bibnamefont {Takagi}},\ and\ \bibinfo {author} {\bibfnamefont {Y.}~\bibnamefont {Shimakawa}},\ }\bibfield  {title} {\bibinfo {title} {Giant multiple caloric effects in charge transition ferrimagnet},\ }\href {https://doi.org/10.1038/s41598-021-91888-8} {\bibfield  {journal} {\bibinfo  {journal} {Sci. Rep.}\ }\textbf {\bibinfo {volume} {11}},\ \bibinfo {pages} {12682} (\bibinfo {year} {2021}{\natexlab{b}})}\BibitemShut {NoStop}%
\bibitem [{\citenamefont {Chen}\ \emph {et~al.}(2023)\citenamefont {Chen}, \citenamefont {Kosugi}, \citenamefont {Goto},\ and\ \citenamefont {Shimakawa}}]{Chen2023a}%
  \BibitemOpen
  \bibfield  {author} {\bibinfo {author} {\bibfnamefont {C.}~\bibnamefont {Chen}}, \bibinfo {author} {\bibfnamefont {Y.}~\bibnamefont {Kosugi}}, \bibinfo {author} {\bibfnamefont {M.}~\bibnamefont {Goto}},\ and\ \bibinfo {author} {\bibfnamefont {Y.}~\bibnamefont {Shimakawa}},\ }\bibfield  {title} {\bibinfo {title} {{Thermal properties and phase transition behaviors of possible caloric materials Bi$_{0.95}$Ln$_{0.05}$NiO$_3$}},\ }\href {http://dx.doi.org/10.1039/D3TA01259J} {\bibfield  {journal} {\bibinfo  {journal} {J. Mater. Chem. A}\ }\textbf {\bibinfo {volume} {11}},\ \bibinfo {pages} {15389} (\bibinfo {year} {2023})}\BibitemShut {NoStop}%
\bibitem [{\citenamefont {Boldrin}(2021)}]{Boldrin2021a}%
  \BibitemOpen
  \bibfield  {author} {\bibinfo {author} {\bibfnamefont {D.}~\bibnamefont {Boldrin}},\ }\bibfield  {title} {\bibinfo {title} {Fantastic barocalorics and where to find them},\ }\href {https://doi.org/10.1063/5.0046416} {\bibfield  {journal} {\bibinfo  {journal} {Appl. Phys. Lett.}\ }\textbf {\bibinfo {volume} {118}},\ \bibinfo {pages} {170502} (\bibinfo {year} {2021})}\BibitemShut {NoStop}%
\bibitem [{\citenamefont {Lloveras}\ and\ \citenamefont {Tamarit}(2023)}]{Lloveras2023a}%
  \BibitemOpen
  \bibfield  {author} {\bibinfo {author} {\bibfnamefont {P.}~\bibnamefont {Lloveras}}\ and\ \bibinfo {author} {\bibfnamefont {J.-L.}\ \bibnamefont {Tamarit}},\ }\bibfield  {title} {\bibinfo {title} {{Chapter 1: Introduction to the barocaloric effect}},\ }in\ \href {https://doi.org/10.1088/978-0-7503-4690-0ch1} {\emph {\bibinfo {booktitle} {Barocaloric Effects in the Solid State: Materials and Methods}}},\ \bibinfo {editor} {edited by\ \bibinfo {editor} {\bibfnamefont {P.}~\bibnamefont {Lloveras}}}\ (\bibinfo  {publisher} {IOP Publishing},\ \bibinfo {address} {Bristol, UK},\ \bibinfo {year} {2023})\ pp.\ \bibinfo {pages} {1--1 to 1--35}\BibitemShut {NoStop}%
\bibitem [{\citenamefont {Lloveras}\ and\ \citenamefont {Tamarit}(2021)}]{Lloveras2021a}%
  \BibitemOpen
  \bibfield  {author} {\bibinfo {author} {\bibfnamefont {P.}~\bibnamefont {Lloveras}}\ and\ \bibinfo {author} {\bibfnamefont {J.-L.}\ \bibnamefont {Tamarit}},\ }\bibfield  {title} {\bibinfo {title} {{Advances and obstacles in pressure-driven solid-state cooling: A review of barocaloric materials}},\ }\href {https://doi.org/10.1557/s43581-020-00002-4} {\bibfield  {journal} {\bibinfo  {journal} {MRS Energy \& Sustainability}\ }\textbf {\bibinfo {volume} {8}},\ \bibinfo {pages} {3} (\bibinfo {year} {2021})}\BibitemShut {NoStop}%
\bibitem [{\citenamefont {Rezaei}\ \emph {et~al.}(2014)\citenamefont {Rezaei}, \citenamefont {Hansmann}, \citenamefont {Bahramy},\ and\ \citenamefont {Arita}}]{Rezaei2014a}%
  \BibitemOpen
  \bibfield  {author} {\bibinfo {author} {\bibfnamefont {N.}~\bibnamefont {Rezaei}}, \bibinfo {author} {\bibfnamefont {P.}~\bibnamefont {Hansmann}}, \bibinfo {author} {\bibfnamefont {M.~S.}\ \bibnamefont {Bahramy}},\ and\ \bibinfo {author} {\bibfnamefont {R.}~\bibnamefont {Arita}},\ }\bibfield  {title} {\bibinfo {title} {{Mechanism of charge transfer/disproportionation in LnCu$_3$Fe$_4$O$_{12}$ (Ln = lanthanides)}},\ }\href {https://doi.org/10.1103/PhysRevB.89.125125} {\bibfield  {journal} {\bibinfo  {journal} {Phys. Rev. B}\ }\textbf {\bibinfo {volume} {89}},\ \bibinfo {pages} {125125} (\bibinfo {year} {2014})}\BibitemShut {NoStop}%
\bibitem [{\citenamefont {Meng}\ \emph {et~al.}(2017)\citenamefont {Meng}, \citenamefont {Zhang}, \citenamefont {Yao}, \citenamefont {Zhang}, \citenamefont {Zhang}, \citenamefont {Liu}, \citenamefont {Meng},\ and\ \citenamefont {Zhang}}]{Meng2017a}%
  \BibitemOpen
  \bibfield  {author} {\bibinfo {author} {\bibfnamefont {J.}~\bibnamefont {Meng}}, \bibinfo {author} {\bibfnamefont {L.}~\bibnamefont {Zhang}}, \bibinfo {author} {\bibfnamefont {F.}~\bibnamefont {Yao}}, \bibinfo {author} {\bibfnamefont {X.}~\bibnamefont {Zhang}}, \bibinfo {author} {\bibfnamefont {W.}~\bibnamefont {Zhang}}, \bibinfo {author} {\bibfnamefont {X.}~\bibnamefont {Liu}}, \bibinfo {author} {\bibfnamefont {J.}~\bibnamefont {Meng}},\ and\ \bibinfo {author} {\bibfnamefont {H.}~\bibnamefont {Zhang}},\ }\bibfield  {title} {\bibinfo {title} {{Theoretical study on the negative thermal expansion perovskite LaCu$_3$Fe$_4$O$_{12}$: pressure-triggered transition of magnetism, charge, and spin state}},\ }\href {https://doi.org/10.1021/acs.inorgchem.7b00458} {\bibfield  {journal} {\bibinfo  {journal} {Inorg. Chem.}\ }\textbf {\bibinfo {volume} {56}},\ \bibinfo {pages} {6371} (\bibinfo {year} {2017})}\BibitemShut {NoStop}%
\bibitem [{\citenamefont {Allub}\ and\ \citenamefont {Alascio}(2012)}]{Allub2012a}%
  \BibitemOpen
  \bibfield  {author} {\bibinfo {author} {\bibfnamefont {R.}~\bibnamefont {Allub}}\ and\ \bibinfo {author} {\bibfnamefont {B.}~\bibnamefont {Alascio}},\ }\bibfield  {title} {\bibinfo {title} {{A thermodynamic model for the simultaneous charge/spin order transition in LaCu$_3$Fe$_4$O$_{12}$}},\ }\href {https://doi.org/10.1088/0953-8984/24/49/495601} {\bibfield  {journal} {\bibinfo  {journal} {J. Phys.: Cond. Matt.}\ }\textbf {\bibinfo {volume} {24}},\ \bibinfo {pages} {495601} (\bibinfo {year} {2012})}\BibitemShut {NoStop}%
\bibitem [{\citenamefont {Azzolina}\ \emph {et~al.}(2020)\citenamefont {Azzolina}, \citenamefont {Bertoni}, \citenamefont {Ecolivet}, \citenamefont {Tokoro}, \citenamefont {Ohkoshi},\ and\ \citenamefont {Collet}}]{Azzolina2020a}%
  \BibitemOpen
  \bibfield  {author} {\bibinfo {author} {\bibfnamefont {G.}~\bibnamefont {Azzolina}}, \bibinfo {author} {\bibfnamefont {R.}~\bibnamefont {Bertoni}}, \bibinfo {author} {\bibfnamefont {C.}~\bibnamefont {Ecolivet}}, \bibinfo {author} {\bibfnamefont {H.}~\bibnamefont {Tokoro}}, \bibinfo {author} {\bibfnamefont {S.~I.}\ \bibnamefont {Ohkoshi}},\ and\ \bibinfo {author} {\bibfnamefont {E.}~\bibnamefont {Collet}},\ }\bibfield  {title} {\bibinfo {title} {Landau theory for non-symmetry-breaking electronic instability coupled to symmetry-breaking order parameter applied to prussian blue analog},\ }\href {https://doi.org/10.1103/PhysRevB.102.134104} {\bibfield  {journal} {\bibinfo  {journal} {Phys. Rev. B}\ }\textbf {\bibinfo {volume} {102}} (\bibinfo {year} {2020})}\BibitemShut {NoStop}%
\bibitem [{\citenamefont {Okada}(1969)}]{Okada1969a}%
  \BibitemOpen
  \bibfield  {author} {\bibinfo {author} {\bibfnamefont {K.}~\bibnamefont {Okada}},\ }\bibfield  {title} {\bibinfo {title} {{Phenomenological theory of antiferroelectric transition. I. Second-order transition}},\ }\href {https://doi.org/10.1143/JPSJ.27.420} {\bibfield  {journal} {\bibinfo  {journal} {J. Phys. Soc. Jpn.}\ }\textbf {\bibinfo {volume} {27}},\ \bibinfo {pages} {420} (\bibinfo {year} {1969})}\BibitemShut {NoStop}%
\bibitem [{\citenamefont {Landau}\ and\ \citenamefont {Lifshitz}(1998)}]{LandauElasticity}%
  \BibitemOpen
  \bibfield  {author} {\bibinfo {author} {\bibfnamefont {L.~D.}\ \bibnamefont {Landau}}\ and\ \bibinfo {author} {\bibfnamefont {A.~P.}\ \bibnamefont {Lifshitz}},\ }\href {https://www.sciencedirect.com/book/9780080570693/theory-of-elasticity} {\emph {\bibinfo {title} {Theory of Elasticity}}}\ (\bibinfo  {publisher} {Elsevier},\ \bibinfo {year} {1998})\BibitemShut {NoStop}%
\bibitem [{\citenamefont {Corrales-Salazar}\ \emph {et~al.}(2017)\citenamefont {Corrales-Salazar}, \citenamefont {Brierley}, \citenamefont {Littlewood},\ and\ \citenamefont {Guzm\'an-Verri}}]{Corrales2017a}%
  \BibitemOpen
  \bibfield  {author} {\bibinfo {author} {\bibfnamefont {A.}~\bibnamefont {Corrales-Salazar}}, \bibinfo {author} {\bibfnamefont {R.~T.}\ \bibnamefont {Brierley}}, \bibinfo {author} {\bibfnamefont {P.~B.}\ \bibnamefont {Littlewood}},\ and\ \bibinfo {author} {\bibfnamefont {G.~G.}\ \bibnamefont {Guzm\'an-Verri}},\ }\bibfield  {title} {\bibinfo {title} {{Landau theory and giant room-temperature barocaloric effect in MF$_3$ metal trifluorides}},\ }\href {https://doi.org/10.1103/PhysRevMaterials.1.053601} {\bibfield  {journal} {\bibinfo  {journal} {Phys. Rev. Mater.}\ }\textbf {\bibinfo {volume} {1}},\ \bibinfo {pages} {053601} (\bibinfo {year} {2017})}\BibitemShut {NoStop}%
\bibitem [{\citenamefont {Marín-Delgado}\ \emph {et~al.}(2024)\citenamefont {Marín-Delgado}, \citenamefont {Moya},\ and\ \citenamefont {Guzmán-Verri}}]{Marin-Delgado2024a}%
  \BibitemOpen
  \bibfield  {author} {\bibinfo {author} {\bibfnamefont {R.}~\bibnamefont {Marín-Delgado}}, \bibinfo {author} {\bibfnamefont {X.}~\bibnamefont {Moya}},\ and\ \bibinfo {author} {\bibfnamefont {G.~G.}\ \bibnamefont {Guzmán-Verri}},\ }\bibfield  {title} {\bibinfo {title} {Landau theory of barocaloric plastic crystals},\ }\href {https://doi.org/10.1088/2515-7655/ad4590} {\bibfield  {journal} {\bibinfo  {journal} {J. Phys.: En.}\ }\textbf {\bibinfo {volume} {6}},\ \bibinfo {pages} {035003} (\bibinfo {year} {2024})}\BibitemShut {NoStop}%
\bibitem [{\citenamefont {Slonczewski}\ and\ \citenamefont {Thomas}(1970)}]{Slonczewski1970a}%
  \BibitemOpen
  \bibfield  {author} {\bibinfo {author} {\bibfnamefont {J.~C.}\ \bibnamefont {Slonczewski}}\ and\ \bibinfo {author} {\bibfnamefont {H.}~\bibnamefont {Thomas}},\ }\bibfield  {title} {\bibinfo {title} {Interaction of elastic strain with the structural transition of strontium titanate},\ }\href {https://doi.org/10.1103/PhysRevB.1.3599} {\bibfield  {journal} {\bibinfo  {journal} {Phys. Rev. B}\ }\textbf {\bibinfo {volume} {1}},\ \bibinfo {pages} {3599} (\bibinfo {year} {1970})}\BibitemShut {NoStop}%
\bibitem [{\citenamefont {Chen}\ \emph {et~al.}(2010)\citenamefont {Chen}, \citenamefont {Long}, \citenamefont {Saito}, \citenamefont {Attfield},\ and\ \citenamefont {Shimakawa}}]{Chen2010a}%
  \BibitemOpen
  \bibfield  {author} {\bibinfo {author} {\bibfnamefont {W.-t.}\ \bibnamefont {Chen}}, \bibinfo {author} {\bibfnamefont {Y.}~\bibnamefont {Long}}, \bibinfo {author} {\bibfnamefont {T.}~\bibnamefont {Saito}}, \bibinfo {author} {\bibfnamefont {J.~P.}\ \bibnamefont {Attfield}},\ and\ \bibinfo {author} {\bibfnamefont {Y.}~\bibnamefont {Shimakawa}},\ }\bibfield  {title} {\bibinfo {title} {{Charge transfer and antiferromagnetic order in the A-site-ordered perovskite LaCu$_3$Fe$_4$O$_{12}$}},\ }\href {https://doi.org/10.1039/C0JM00767F} {\bibfield  {journal} {\bibinfo  {journal} {J. Mater. Chem.}\ }\textbf {\bibinfo {volume} {20}},\ \bibinfo {pages} {7282} (\bibinfo {year} {2010})}\BibitemShut {NoStop}%
\bibitem [{\citenamefont {Long}\ \emph {et~al.}(2012)\citenamefont {Long}, \citenamefont {Kawakami}, \citenamefont {Chen}, \citenamefont {Saito}, \citenamefont {Watanuki}, \citenamefont {Nakakura}, \citenamefont {Liu}, \citenamefont {Jin},\ and\ \citenamefont {Shimakawa}}]{Long2012a}%
  \BibitemOpen
  \bibfield  {author} {\bibinfo {author} {\bibfnamefont {Y.-w.}\ \bibnamefont {Long}}, \bibinfo {author} {\bibfnamefont {T.}~\bibnamefont {Kawakami}}, \bibinfo {author} {\bibfnamefont {W.-t.}\ \bibnamefont {Chen}}, \bibinfo {author} {\bibfnamefont {T.}~\bibnamefont {Saito}}, \bibinfo {author} {\bibfnamefont {T.}~\bibnamefont {Watanuki}}, \bibinfo {author} {\bibfnamefont {Y.}~\bibnamefont {Nakakura}}, \bibinfo {author} {\bibfnamefont {Q.-q.}\ \bibnamefont {Liu}}, \bibinfo {author} {\bibfnamefont {C.-q.}\ \bibnamefont {Jin}},\ and\ \bibinfo {author} {\bibfnamefont {Y.}~\bibnamefont {Shimakawa}},\ }\bibfield  {title} {\bibinfo {title} {{Pressure effect on intersite charge transfer in A-site-ordered double-perovskite-structure oxide}},\ }\href {https://doi.org/10.1021/cm301267e} {\bibfield  {journal} {\bibinfo  {journal} {Chem. Mater.}\ }\textbf {\bibinfo {volume} {24}},\ \bibinfo {pages} {2235} (\bibinfo {year} {2012})}\BibitemShut {NoStop}%
\bibitem [{\citenamefont {Mizzi}\ and\ \citenamefont {Maiorov}(2024)}]{Mizzi2024a}%
  \BibitemOpen
  \bibfield  {author} {\bibinfo {author} {\bibfnamefont {C.~A.}\ \bibnamefont {Mizzi}}\ and\ \bibinfo {author} {\bibfnamefont {B.}~\bibnamefont {Maiorov}},\ }\bibfield  {title} {\bibinfo {title} {Enabling resonant ultrasound spectroscopy in high magnetic fields},\ }\href {https://doi.org/10.1121/10.0026124} {\bibfield  {journal} {\bibinfo  {journal} {J. Acoust. Soc. Am.}\ }\textbf {\bibinfo {volume} {155}},\ \bibinfo {pages} {3505} (\bibinfo {year} {2024})}\BibitemShut {NoStop}%
\bibitem [{\citenamefont {Azuma}\ \emph {et~al.}(2007)\citenamefont {Azuma}, \citenamefont {Carlsson}, \citenamefont {Rodgers}, \citenamefont {Tucker}, \citenamefont {Tsujimoto}, \citenamefont {Ishiwata}, \citenamefont {Isoda}, \citenamefont {Shimakawa}, \citenamefont {Takano},\ and\ \citenamefont {Attfield}}]{Azuma2007a}%
  \BibitemOpen
  \bibfield  {author} {\bibinfo {author} {\bibfnamefont {M.}~\bibnamefont {Azuma}}, \bibinfo {author} {\bibfnamefont {S.}~\bibnamefont {Carlsson}}, \bibinfo {author} {\bibfnamefont {J.}~\bibnamefont {Rodgers}}, \bibinfo {author} {\bibfnamefont {M.~G.}\ \bibnamefont {Tucker}}, \bibinfo {author} {\bibfnamefont {M.}~\bibnamefont {Tsujimoto}}, \bibinfo {author} {\bibfnamefont {S.}~\bibnamefont {Ishiwata}}, \bibinfo {author} {\bibfnamefont {S.}~\bibnamefont {Isoda}}, \bibinfo {author} {\bibfnamefont {Y.}~\bibnamefont {Shimakawa}}, \bibinfo {author} {\bibfnamefont {M.}~\bibnamefont {Takano}},\ and\ \bibinfo {author} {\bibfnamefont {J.~P.}\ \bibnamefont {Attfield}},\ }\bibfield  {title} {\bibinfo {title} {{Pressure-induced intermetallic valence transition in BiNiO$_3$}},\ }\href {https://doi.org/10.1021/ja074880u} {\bibfield  {journal} {\bibinfo  {journal} {J. Am. Chem. Soc.}\ }\textbf {\bibinfo {volume} {129}},\ \bibinfo {pages} {14433} (\bibinfo {year} {2007})}\BibitemShut {NoStop}%
\bibitem [{\citenamefont {Seda}\ and\ \citenamefont {Hearne}(2004)}]{Takele2004a}%
  \BibitemOpen
  \bibfield  {author} {\bibinfo {author} {\bibfnamefont {T.}~\bibnamefont {Seda}}\ and\ \bibinfo {author} {\bibfnamefont {G.~R.}\ \bibnamefont {Hearne}},\ }\bibfield  {title} {\bibinfo {title} {{Pressure induced Fe$^{2+}$+Ti$^{4+}$ $\to$ Fe$^{3+}$+Ti$^{3+}$+ intervalence charge transfer and the Fe$^{3+}$/Fe$^{2+}$ ratio in natural ilmenite (FeTiO$_3$) minerals}},\ }\href {https://doi.org/10.1088/0953-8984/16/15/021} {\bibfield  {journal} {\bibinfo  {journal} {J. Phys.: Cond. Matt.}\ }\textbf {\bibinfo {volume} {16}},\ \bibinfo {pages} {2707} (\bibinfo {year} {2004})}\BibitemShut {NoStop}%
\bibitem [{\citenamefont {Liu}\ \emph {et~al.}(2020)\citenamefont {Liu}, \citenamefont {Sakai}, \citenamefont {Yang}, \citenamefont {Li}, \citenamefont {Liu}, \citenamefont {Ye}, \citenamefont {Qin}, \citenamefont {Chen}, \citenamefont {Agrestini}, \citenamefont {Chen}, \citenamefont {Liao}, \citenamefont {Haw}, \citenamefont {Baudelet}, \citenamefont {Ishii}, \citenamefont {Nishikubo}, \citenamefont {Ishizaki}, \citenamefont {Yamamoto}, \citenamefont {Pan}, \citenamefont {Fukuda}, \citenamefont {Ohashi}, \citenamefont {Matsuno}, \citenamefont {Machida}, \citenamefont {Watanuki}, \citenamefont {Kawaguchi}, \citenamefont {Arevalo-Lopez}, \citenamefont {Jin}, \citenamefont {Hu}, \citenamefont {Attfield}, \citenamefont {Azuma},\ and\ \citenamefont {Long}}]{Liu2020a}%
  \BibitemOpen
  \bibfield  {author} {\bibinfo {author} {\bibfnamefont {Z.}~\bibnamefont {Liu}}, \bibinfo {author} {\bibfnamefont {Y.}~\bibnamefont {Sakai}}, \bibinfo {author} {\bibfnamefont {J.}~\bibnamefont {Yang}}, \bibinfo {author} {\bibfnamefont {W.}~\bibnamefont {Li}}, \bibinfo {author} {\bibfnamefont {Y.}~\bibnamefont {Liu}}, \bibinfo {author} {\bibfnamefont {X.}~\bibnamefont {Ye}}, \bibinfo {author} {\bibfnamefont {S.}~\bibnamefont {Qin}}, \bibinfo {author} {\bibfnamefont {J.}~\bibnamefont {Chen}}, \bibinfo {author} {\bibfnamefont {S.}~\bibnamefont {Agrestini}}, \bibinfo {author} {\bibfnamefont {K.}~\bibnamefont {Chen}}, \bibinfo {author} {\bibfnamefont {S.-C.}\ \bibnamefont {Liao}}, \bibinfo {author} {\bibfnamefont {S.-C.}\ \bibnamefont {Haw}}, \bibinfo {author} {\bibfnamefont {F.}~\bibnamefont {Baudelet}}, \bibinfo {author} {\bibfnamefont {H.}~\bibnamefont {Ishii}}, \bibinfo {author} {\bibfnamefont {T.}~\bibnamefont {Nishikubo}}, \bibinfo {author} {\bibfnamefont {H.}~\bibnamefont {Ishizaki}}, \bibinfo {author}
  {\bibfnamefont {T.}~\bibnamefont {Yamamoto}}, \bibinfo {author} {\bibfnamefont {Z.}~\bibnamefont {Pan}}, \bibinfo {author} {\bibfnamefont {M.}~\bibnamefont {Fukuda}}, \bibinfo {author} {\bibfnamefont {K.}~\bibnamefont {Ohashi}}, \bibinfo {author} {\bibfnamefont {K.}~\bibnamefont {Matsuno}}, \bibinfo {author} {\bibfnamefont {A.}~\bibnamefont {Machida}}, \bibinfo {author} {\bibfnamefont {T.}~\bibnamefont {Watanuki}}, \bibinfo {author} {\bibfnamefont {S.~I.}\ \bibnamefont {Kawaguchi}}, \bibinfo {author} {\bibfnamefont {A.~M.}\ \bibnamefont {Arevalo-Lopez}}, \bibinfo {author} {\bibfnamefont {C.}~\bibnamefont {Jin}}, \bibinfo {author} {\bibfnamefont {Z.}~\bibnamefont {Hu}}, \bibinfo {author} {\bibfnamefont {J.~P.}\ \bibnamefont {Attfield}}, \bibinfo {author} {\bibfnamefont {M.}~\bibnamefont {Azuma}},\ and\ \bibinfo {author} {\bibfnamefont {Y.}~\bibnamefont {Long}},\ }\bibfield  {title} {\bibinfo {title} {{Sequential spin state transition and intermetallic charge transfer in PbCoO$_3$}},\ }\href
  {https://doi.org/10.1021/jacs.9b13508} {\bibfield  {journal} {\bibinfo  {journal} {J. Am. Chem. Soc.}\ }\textbf {\bibinfo {volume} {142}},\ \bibinfo {pages} {5731} (\bibinfo {year} {2020})}\BibitemShut {NoStop}%
\end{thebibliography}

%

\end{document}